\documentclass[twocolumn]{aastex631} 
\usepackage{hyperref}
\shorttitle{4U 1730--22}
\shortauthors{Li et al.}
\graphicspath{{./}{figures/}}

\begin{document}

\title{Discovery of a $584.65$ Hz Burst Oscillation in the Low-mass X-Ray Binary 4U 1730--22}

\author{Zhaosheng Li}
\affiliation{Key Laboratory of Stars and Interstellar Medium, Xiangtan University, Xiangtan 411105, Hunan, P.R. China}

\correspondingauthor{Zhaosheng Li, Yuanyue Pan}
\email{lizhaosheng@xtu.edu.cn, panyy@xtu.edu.cn}

\author{Wenhui Yu}
\affiliation{Key Laboratory of Stars and Interstellar Medium, Xiangtan University, Xiangtan 411105, Hunan, P.R. China}

\author{Yongqi Lu}
\affiliation{Key Laboratory of Stars and Interstellar Medium, Xiangtan University, Xiangtan 411105, Hunan, P.R. China}

\author{Yuanyue Pan}
\affiliation{Key Laboratory of Stars and Interstellar Medium, Xiangtan University, Xiangtan 411105, Hunan, P.R. China}

\author{Maurizio Falanga}
\affiliation{International Space Science Institute (ISSI), Hallerstrasse 6, 3012 Bern, Switzerland
}
\affiliation{Physikalisches Institut, University of Bern, Gesellsschaftstrasse 6, 3012 Bern, Switzerland
}

\begin{abstract}
Type-I X-ray burst oscillations are powered by thermonuclear energy released on the neutron star (NS) surface in low-mass X-ray binaries (LMXBs), where the burst oscillation frequencies are close to the NS spin rates.
In this work, we report the detection of oscillation at 584.65 Hz during the cooling tail of a type-I X-ray bursts observed from the accreting NS LMXB 4U~1730--22 on 2022 March 20, by the Neutron star Interior Composition Explorer telescope. The oscillation signal showed a strong Leahy power, $P_{\rm m}\sim54.04$, around 584.65 Hz, which has single-trial and multiple-trials confidence levels of $7.05\sigma$ and  $4.73\sigma$, respectively. The folded pulse profile of the oscillation in the 0.2--10 keV band showed a sinusoidal shape with the fractional rms amplitude  of $(8.0\pm1.1)\%$. We found  the oscillation frequency showed insignificant upward drifting, i.e., less than 0.3 Hz, during the cooling tail, similar to the behavior appearing in accreting millisecond X-ray pulsars (AMXP), and indicate the source could be an AMXP spinning at 1.71 ms.  

\end{abstract}



\section{Introduction} \label{sec:intro}

The X-ray source 4U 1730--22 was discovered by Uhuru during its  X-ray outburst in 1972 \citep{Cominsky78,Forman78}.  The quiescent state of the source has been observed by Chandra, which associates with the faint X-ray source CXOU J173357.5--220156 \citep{Tomsick07}. Based on the persistent and quiescent X-ray spectra, 4U 1730--22 was classified as a probable neutron star low-mass X-ray binary \citep[NS LMXB; ][]{Tanaka96,Chen97}. On 2021 June 7, the MAXI/GSC team reported the activities of an X-ray transient, which is associated with 4U 1730--22 after 50 yr of being at a quiescent state as confirmed by Swift follow-up observations \citep{ ATel14686,ATel14688,ATel14683,ATel14757}. The position of 4U 1730--22 has been localized in X-ray and optical wavelengths \citep{ATel14688,ATel14693}, which is consistent with the localization by Chandra  \citep{Tomsick07}.  


Type-I X-ray bursts are triggered from unstable thermonuclear burning of accreting material on the NS surface \citep[][]{Lewin93,Galloway08,Galloway21}.  The first type-I X-ray burst from the source was detected by
Neutron star Interior Composition Explorer  \citep[NICER;][]{Atel14769}, confirming definitively, that the source is a transient LMXB hosting an accreting NS. Moreover, the optical counterpart of 4U 1730--22 has been identified, which shows strong hydrogen and weaker helium emission lines indicating a companion star in the main sequence \citep{ATel14693,ATel14694}.
 
Besides type-I X-ray bursts, the detection of coherent pulsation can also confirm the compact object in an LMXB as an NS. In addition, measuring the NS spin periods are also important in may other aspects, such as studying the spin evolution during the accretion process, evaluating the broadening effects of spectral lines on the NS surface \citep{Chang06}, measuring the orbit parameters \citep[see, e.g.,][]{Strohmayer18}, determining the NS masses and radii from photospheric radius expansion  bursts \citep{Suleimanov20}, and constraining on the NS equation of state especially from ultrafast rotational NS \citep[i.e., submillisecond pulsars,][]{Haensel09}. The spin periods of a large fraction of NSs have been measured from their coherent emissions observed in radio or X-ray bands or both \citep[see e.g.,][]{Manchester05,Liu06,Liu07,Walter15,Patruno21}. 

During type-I X-ray  bursts, an NS may produce inhomogeneous thermal emissions on the surface, which manifests as burst oscillations \citep{Strohmayer96}.  The type-I X-ray burst oscillations provide us with an indirect method to measure the NS spin periods, due to the fact that the oscillation frequencies measured in accreting millisecond X-ray pulsars (AMXPs) are very close to their coherent spin frequencies within a few hertz during the persistent emissions (e.g., \citealt{Chakrabarty03}; see \citealt{Watts12} and \citealt{Bhattacharyya22} for reviews). Totally, five persistent  AMXPs have been observed burst oscillations, and their oscillation frequencies usually showed negligible upward drifting during the cooling tails \citep{Chakrabarty03,Galloway08,Bilous19}. 


NICER was launched and installed on the International Space Station on 2017 June 3,  which has the capabilities to collect X-ray photons in the energy of 0.2--12 keV with absolute time resolution as high as 100 ns \citep{gendreau2017searching}. After 5 yr of operation, NICER has been observed $\sim56\%$ of all known NS LMXBs  in the MINBAR catalog (\citealt{Galloway20}; see also the webpage \footnote{\url{https://personal.sron.nl/~jeanz/bursterlist.html}}) that have exhibited type-I X-ray bursts, including 23 sources with the detection of burst oscillation (or candidates) previously by RXTE and Swift \citep[see][and references therein]{Bilous19}.  Benefiting from its high timing accuracy and relatively large collecting area,  NICER has detected burst oscillations from 4U 1728--34 at $\sim$362.5 Hz \citep{Mahmoodifar19} and SAX J1808.4--3658 at 401 Hz \citep{Bult19}, which were previously found by \textit{RXTE} \citep{Strohmayer96,Chakrabarty03}. Recently, the burst oscillation candidate from XTE J1739--285  has been observed at 386.5 Hz \citep{Bult21b}, rather than 1122 Hz reported by \citet{Kaaret07}. However, most of the sources have not reported the detection of burst oscillations \citep[see, e.g., Aql X--1 and 4U 1636--536;][]{Li21,Guver22,Zhao22}.

In this work, we report the detection of the burst oscillation at the frequency $\sim584.65$ Hz from 4U 1730--22.

\section{Observations} \label{sec:obs} 

\begin{figure*}
\includegraphics[width=9cm]{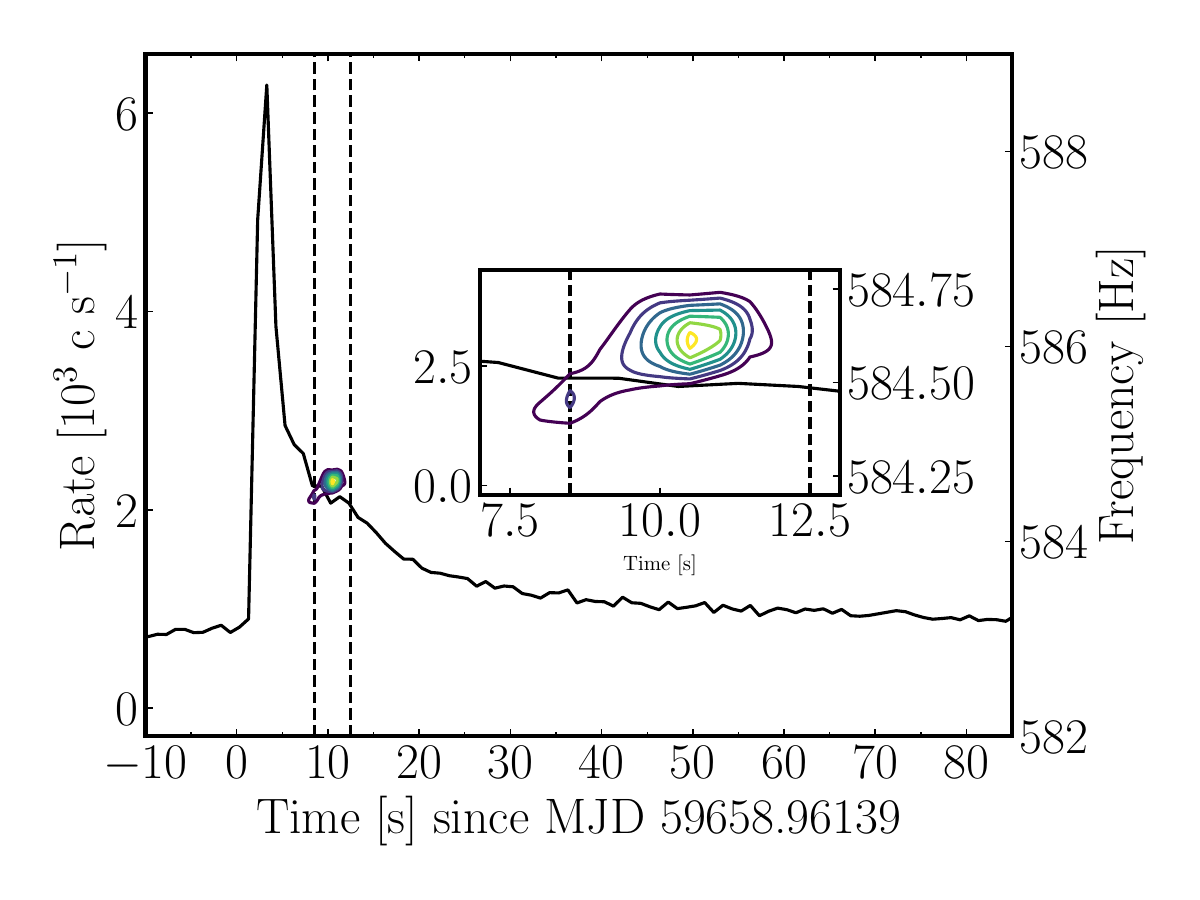}
\includegraphics[width=9cm]{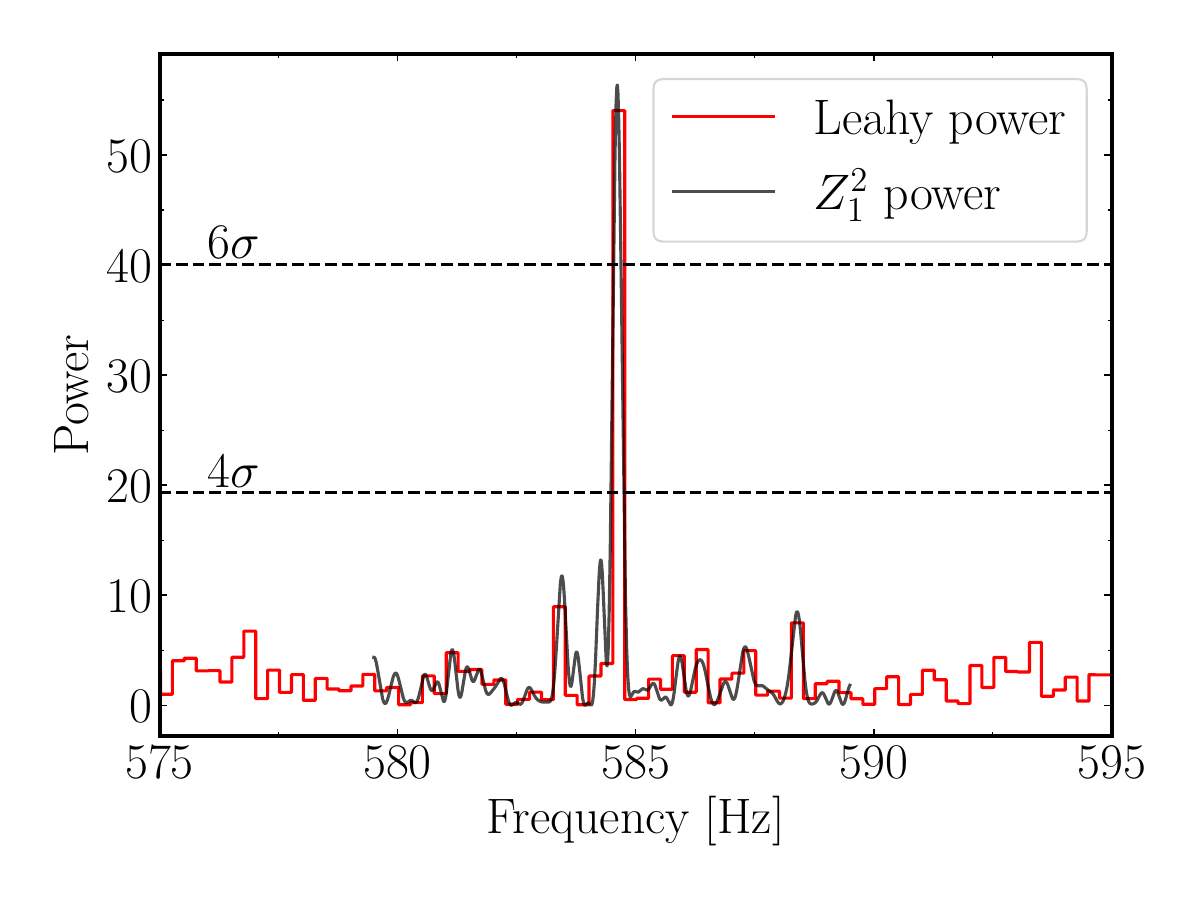}
\caption{Left panel: the 1 s binned light curve of the burst from ObsID 5202200113 (burst \#4). We search the time interval between $-5$ and 75 s. The contours mark the $Z_1^2$ power from 25 to 55 with the steps of 5. The insert zoomed-in panel shows the time interval  where the oscillation was detected. Right panel, the Leahy and $Z_1^2$ power marked as vertical dashed lines, during the burst interval $t\sim8.5-12.5$ s, in the left panel. The horizontal dashed lines represent the single-trial significance of $4\sigma$ and $6\sigma$, respectively. }
\label{fig:pow} 
\end{figure*}

In this work, we analyzed all public archived NICER observations of 4U 1730--22 on Modified Julian Date (MJD) 59642.0--59775.0 that have a total unfiltered exposure of 447.37 ks. The observation IDs (ObsIDs) include 42022001$aa$, 52022001$bb$, and  46390101$cc$, where $aa$, $bb$ and $cc$ run from 01 to 43, 01 to 19, and 01 to 83, respectively. We processed the NICER data analysis using HEASOFT V6.30.1 and the NICER Data Analysis Software (NICERDAS). We adopted the default selection criteria by using \texttt{nicerl2}  to filter the cleaned event data. We then applied barycentric corrections using the tool \texttt{barycorr}  to all the event data  employing the source coordinates $\rm{R.A.=263^\circ.489792}$,  $\rm{decl.=-22^\circ.032472}$ \citep{Tomsick07} and the JPL-DE430 ephemeris. The 1 s light curves in the energy of 0.5--10 keV of all observations were extracted. From the light curves, a total of  16 type-I X-ray bursts have been identified. The observation log and properties of all bursts are listed in Table~\ref{table:data}.

\begin{figure}
\includegraphics[width=8.5cm]{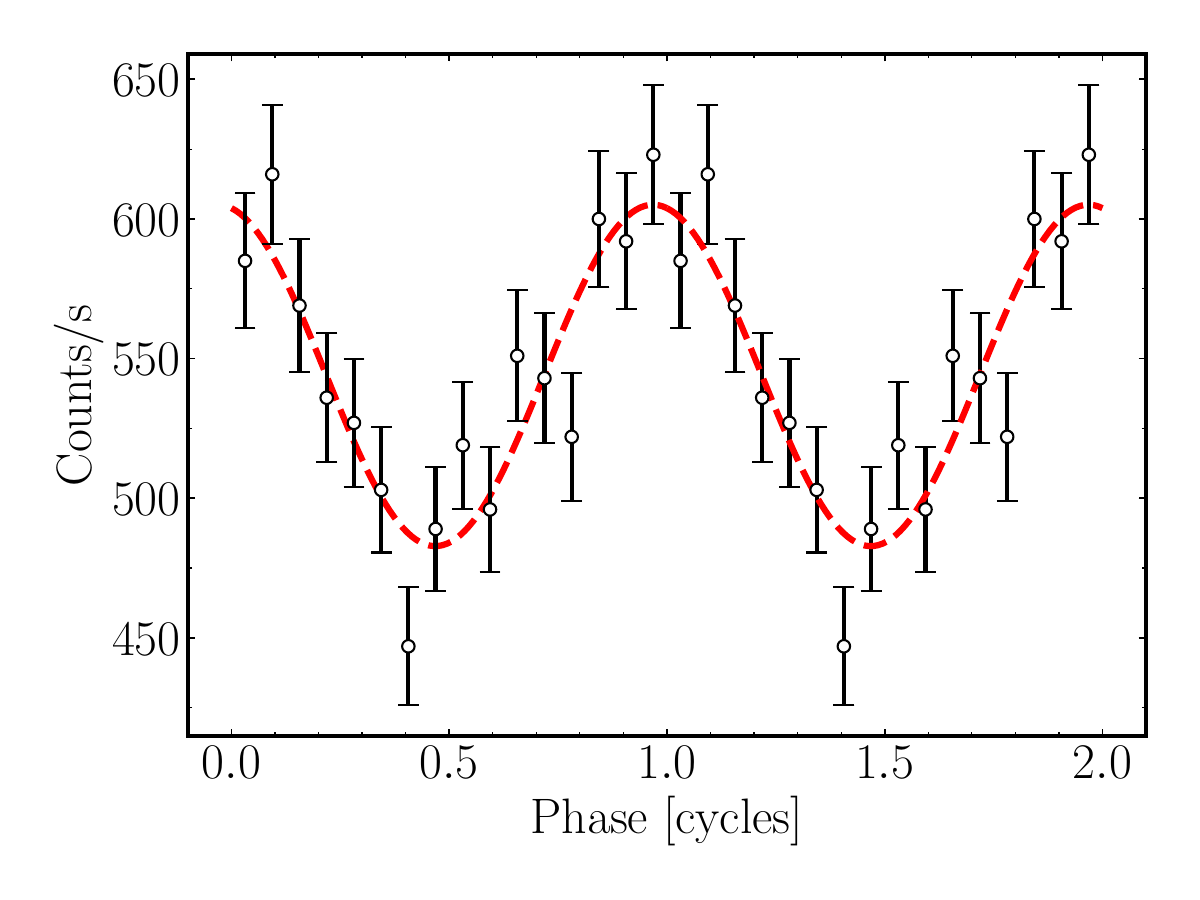}
\caption{The folded pulse profile  of the burst oscillation in the energy range of 0.2--10 keV during the interval marked as vertical dashed lines in the left panel of Fig.~\ref{fig:pow}. The red dashed line represents the best fitted sinusoidal model, $A+B\sin(2\pi\nu t-\phi_0)$. In order to clarity, two cycles are shown. }
\label{fig:prof} 
\end{figure}

\section{Results} \label{sec:results}

\subsection{Burst Oscillation}\label{sec:fft_search}

Since the spin period or burst oscillation was previously unknown in 4U 1730--22, we carried out blind searches of all type-I X-ray bursts in Table~\ref{table:data} for testing the presence of coherent burst oscillations, where the time interval of each burst starts $-5$ s prior to the burst trigger. To preserve as many photons as possible, the cleaned event files in the 0.2--10 keV energy range were used.
We applied the fast Fourier transform (FFT) technique with Leahy normalization to calculate the power spectra.  For a pure Poisson counting process without any deterministic signal, the Leahy power satisfies a $\chi^2$ distribution with two degrees of freedom \citep{Leahy83}. We adopt the slide window of $\Delta T=4$ s, each new window moving forward 0.5 s  with respect to the
previous one.  For each window, the statistically independent Fourier frequencies between 50 and 2000 Hz with steps of 0.25 Hz \footnote{In practice, however, if the duration between the first and last photons in a 4 s window is slightly smaller than 4 s, then the Fourier frequency bin is slightly larger than 0.25 Hz. It has a negligible effect for estimating the detection significance of burst oscillation. } and the corresponding power were recorded \citep[see, e.g.,][]{Bilous19, Bult21b}.  


We found the cooling tail in burst \#4 shows a strong oscillation signal at $\sim 584.65$ Hz with the Leahy power peaked at $P_{\rm m}=54.04$, well exceeding $P_{\rm m}=2$ expected from white noise. It corresponds to a probability of $1.84\times10^{-12}$ produced by chance for a single trial, which converts to the confidence level of $7.05\sigma$. If considering all the searched frequencies ($N_{\rm Freq}=1950\times4$) and segments ($N_{\rm seg} =160$), the total trials are $N=N_{\rm Freq}\times N_{\rm seg}=1.248\times10^6$. It should be noticed that the segments are highly overlapped and not independent. We conservatively estimated that the confidence level of multiple trials is $2.30\times10^{-6}$, i.e., $4.73\sigma$. Finally, further adjusting the trial count to include all 16 bursts, we obtained $3.68\times10^{-5}$ and $4.13\sigma$.

In order to search for the burst oscillations more accurately, we applied the $Z_1^2$-test statistics, in which the $Z_1^2$ power also satisfies the $\chi^2$ distribution with two degrees of freedom,  based on {\tt Stingray} \citep{Buccheri83, Huppenkothen19} in the frequency range between 579.5--589.5 Hz with steps of 0.01 Hz. The cleaned event files in the 0.2--10 keV energy range are also used by applying a moving window method with the same window size and step as the above-mentioned FFT method. We also reproduced the oscillation signal for the burst oscillation around 584.61 Hz, in which the $Z_1^2$ power peaked at 56.37. It converts to a probability of $5.75\times10^{-13}$ produced by a random process for a single trial, that is, $7.21\sigma$. If considering the total searched frequencies ($N_{\rm Freq}=1000$) and segments (same as the FFT method), the probability of multiple trials is $9.19\times10^{-8}$. Again, the segments are treated as independent and accounting for all searched bursts, the significance of the burst oscillation is modestly estimated as $4.82\sigma$.



In the left panel of Fig.~\ref{fig:pow}, we produced the light curve of burst \#4, and showed the contours of $Z_1^2$ with the power changing from 25 to 55. We found the  upward drifting of the oscillation frequency smaller than 0.3 Hz.

The   fractional rms (root-mean-square)  amplitude is  calculated from the relation,
\begin{equation}
    A_{rms}=\left(\frac{P_{\rm s}}{N_{\rm m}}\right)^{1/2}\frac{N_{\rm m}}{N_{\rm m}-N_{\rm bkg}},
\end{equation}
where $P_{\rm s}$ is the power of the signal, $N_{\rm m}$ and $N_{\rm bkg}$ are the total and background X-ray photons collected during the burst interval. Compared with the high count rate of the burst, the background contribution can be neglected \citep[see also][]{Mahmoodifar19,Remillard22}. We adopted an approximate relation, $A_{rms}\approx\sqrt{P_{\rm s}/N_{\rm m}}$.  The power $P_{\rm s}$ and its standard deviation were estimate from \citet{Groth75} and \citet{Vaughan94} \citep[see also][]{Bilous19} based on the measured power, $P_{\rm m}$. We obtained the  fractional rms amplitude  to be $(8.0\pm1.1)\%$.

Using the oscillation frequency at the highest $Z_1^2$ power, in Fig.~\ref{fig:prof}, we folded the pulse profile of burst \#4 in the 0.2--10 keV in 16 phase bins from the 4 s interval marked by vertical dashed lines in the left panel in Fig.~\ref{fig:pow}. The sinusoidal model, $A+B\sin(2\pi\nu t-\phi_0)$, has been applied to fit the pulse profile, resulting in the best-fitted values and 1$\sigma$ uncertainties,  $A=544.0\pm5.8$, $B=61.1\pm8.2$ and $\phi_0=-(0.56\pm0.02)\pi$, respectively. The goodness of fit has the minimum $\chi^2$ of 14.2 for 13 degrees of freedom, i.e., 16 phase bins minus three model parameters, indicating a good fit. The fractional rms amplitude from the folded profile, $B/\sqrt{2}A$, is consistent with the previous result.

We have also searched for the possible presence of the second harmonics related to the burst oscillation frequency. We have found no significant signal and put an upper limits of the fractional rms amplitude of 2.6\% and the significance level of $1.3\sigma$ for a single trial. 

\subsection{X-ray Burst light curve simulations}

\begin{figure}
\includegraphics[width=8.5cm]{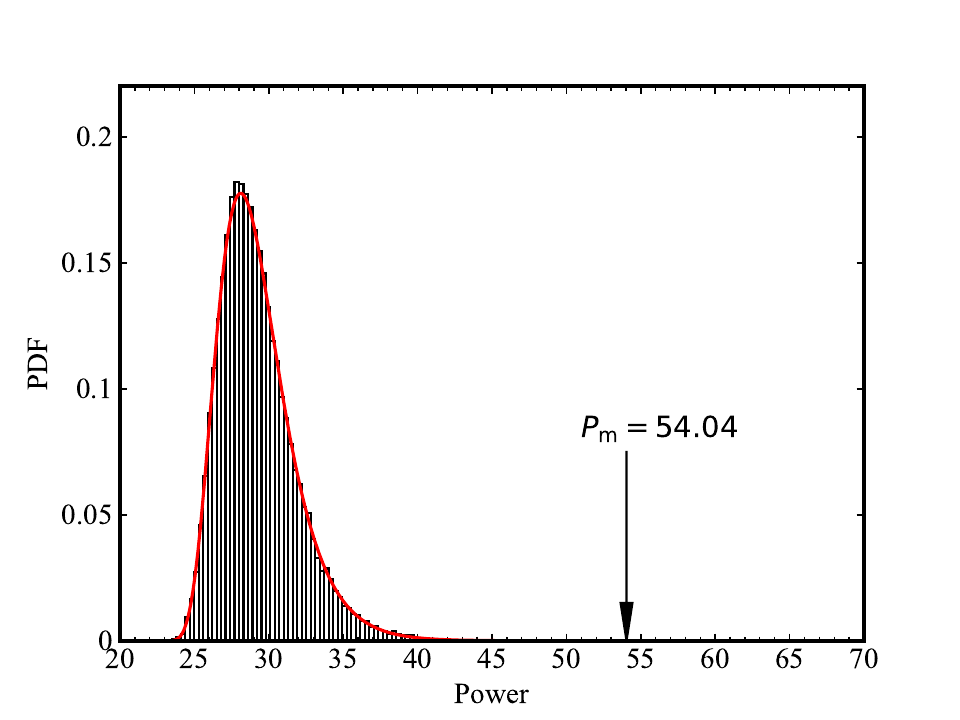}
\caption{ The distribution of simulating FFT powers. The red curve shows a lognormal fit. The arrow marks the measured FFT power from the burst \#4. }
\label{fig:simu} 
\end{figure}

The significance of the burst oscillation is calculated from overlapped timing windows, which are statically independent. In order to estimate the probability of detecting the oscillation signal by chance, we performed a sophisticated method through numerical simulations to take into account the effectiveness of the nonstationary of the burst light curve and the skewing effects of using overlapping windows.  We constructed a sample of artificial light curves by using a method similar to \citet{Bilous19} and \citet{Bult21}; see also the webpage.\footnote{\url{https://jmichaelburgess.com/lc/}} For each simulated light curve, we applied exactly the same searching procedure as mentioned in Sec.~\ref{sec:fft_search}, and then, the maximum FFT power in the frequency range between 50 and 2000 Hz is preserved. After running $5\times10^4$ simulations, we found that none of the simulated maximum powers were higher than $P_{\rm m}$.  We also found that the maximum powers satisfy a lognormal distribution (see Fig~\ref{fig:simu}). The probability to obtain the power exceeding the measured value, $P_{\rm m}=54.04$, is $2.67\times 10^{-6}$.  This shows directly that the detected burst oscillation is unlikely to occur by random chance.

\section{Discussion} \label{sec:diss}

We detected 16 type-I X-ray bursts from 4U 1730--22 by analyzing its 2021 and 2022 NICER observations. For the first time, we found that one burst in 4U 1730--22 appears as a strong burst oscillation signal during the cooling tail. The maximum Leahy power is 54.04, corresponding to the significance of $7.05\sigma$ for a single trial, $4.73\sigma$ for multiple trials, and $4.13\sigma$ with all 16 bursts included, which are more significant than the burst oscillation powers from 4U 1728--34 \citep{Mahmoodifar19} and XTE J1739--285 \citep{Bult21} observed by NICER.  Moreover, the single burst oscillation signal for 4U 1730--22 is stronger than detections from SAX J1750.8--2900, HETE J1900.1-2455 \citep{Bilous19}, and 4U 0614+09 \citep{Strohmayer08}, but weaker than  SAX J1810.8--2609 \citep{Bilous18}. The folded pulsation of the burst oscillation showed a sinusoidal shape, with a fractional rms amplitude of $(8.0\pm1.1)\%$. 
Both features are typical characteristics shown in many other NS LMXBs \citep[see, e.g.,][]{Galloway08}. The burst decay oscillation and the fractional rms amplitude shown in 4U 1730--22 can be explained by the surface mode model, where the oscillation waves are excited at the burst raise in the NS ocean and spread for a large fraction of the cooling tail, or the asymmetric cooling wake model, where different parts of the NS surface cool at different rates \citep[see, e.g.,][]{Heyl04,Watts12,Mahmoodifar16,Bhattacharyya22}.


The oscillation frequency is around 584.65 Hz, which indicates a fast rotational NS in 4U 1730--22 with a spin period of around 1.71 ms. Moreover, the oscillation frequency showed small upward drifting, less than 0.3 Hz,  during the cooling tail, similar to the behaviors appearing in AMXPs \citep[see, e.g., ][]{Chakrabarty03,Watts12, Bult19}. Hence, we suggest that 4U 1730--22 could be also an  AMXP spinning at 1.71 ms.  During the process of this work, 4U 1730--22 is still in outburst, and more data will be collected. Our observed burst oscillation can be searched and verified from possible new triggered type-I X-ray bursts from this source in near future. The spectral studies of all type-I X-ray bursts will be published elsewhere. With an almost known spin period, it is helpful to perform exhaustive searching for the coherent pulsation from the persistent emissions with reduced parameter spaces when the outburst stops \citep[e.g.,][]{Strohmayer18}. If the spin frequency is confirmed, 4U 1730--22 would be one of the fastest accreting NSs, which belongs to the subpopulation of LMXBs with a high spin frequency centered at $\approx575$ Hz \citep{Patruno17}. Moreover, we note that the effective surface temperature of 4U 1730--22 during its quiescent state was hot, that is, $1.51\times 10^6$ K \citep{Tomsick07}.  Rather than a coincidence, a rapidly rotating NS tends to excite an $r$-mode instability to balance the spin-up torque, resulting in reheating the star \citep{Gusakov14,Gusakov14b,Patruno17}.

In addition, radio observations are highly encouraged to carried out, to search for the coherent radio pulsation and catch the possible  transition from accretion-powered to rotation-powered states, when the source evolves into the X-ray quiescent state \citep[see][]{Papitto22}.

\begin{table*}
\begin{center} 
\caption{Summary of burst observations.  \label{table:data}}
\begin{tabular}{ccccc} 
\hline\\ 
{\centering Burst No. } &
{\centering NICER} &
{\centering  Burst Trigger } &
{\centering  ObsID Start Time} &
{\centering  Peak Count Rate$^{\rm{a}}$}  \\
 & ObsID & (TDB MJD) & (YYYY-MM-DD)  & ($10^3$\,ct\,s$^{-1}$) 
\\ [0.01cm]
\hline\\
1 & 4202200125  & 59404.55780  &  2021-07-09  & 3.14 \\
2 & 5202200101  & 59639.33493 &   2022-03-01  & 3.67 \\
3 & 5202200112  & 59657.91327 &   2022-03-19  & 6.31 \\
4 & 5202200113  & 59658.96139 &   2022-03-20  & 6.29\\
5 & 4639010102  & 59664.12261 &  2022-03-26   & 2.71 \\
6 & 4639010104  & 59666.95125 &   2022-03-28  & 2.98 \\
7 & 4639010113  & 59675.59719 &   2022-04-06  & 2.32 \\
8 & 4639010116  & 59678.77329 &   2022-04-09  & 6.99 \\
9 & 4639010131  & 59695.09694 &   2022-04-26  & 7.93\\
10 &4639010141 & 59718.3307& 2022-05-19 & 5.45 \\
11 &4639010146 &  59723.8250 & 2022-05-24 & 2.78 \\
12 &4639010160 &  59739.4289 & 2022-06-09 & 5.79 \\
13 &4639010160 & 59739.8745 & 2022-06-09 & 6.87 \\
14 &4639010166 &  59747.6837 & 2022-06-17 & 5.91 \\
15 &4639010175 &  59756.8607 &2022-06-26  & 5.83 \\
16 &4639010179 &  59762.7372 &2022-07-01  & 5.59 \\

\hline 
\end{tabular} 

$^{\mathrm{a}}$ The peak count rates are measured from the 1 s cleaned light curves in the energy range of 0.5--10 keV. 

\end{center}
\end{table*}

\begin{acknowledgments}

The authors thank the referee for valuable comments that improved our manuscript. Z.L. and Y.Y.P. were supported by National Natural Science Foundation of China (12130342, U1938107). This research has made use of data obtained from the High Energy Astrophysics Science Archive Research Center (HEASARC), provided by NASA’s Goddard Space Flight Center.

\end{acknowledgments}

%

\vspace{5mm}
\facilities{NICER \citep{Gendreau12}.}


\software{Stingray \citep{Huppenkothen19}, HEASOFT \citep{HEAsoft}.
          }



\bibliography{burst}{}
\bibliographystyle{aasjournal}



\end{document}